\documentstyle[12pt,epsfig]{article}
\def\lapp{{\ \lower 0.6ex \hbox{$\buildrel<\over\sim$}\ }}
\def\gapp{{\ \lower 0.6ex \hbox{$\buildrel>\over\sim$}\ }}
\begin{document}
\begin{titlepage}
\vspace*{-1cm}
\begin{flushright}
DTP/95/70 \\
August 1995 \\
\end{flushright}
\vskip 1.cm
\begin{center}
{\Large\bf Hidden Top Quark Decay to Charged Higgs Scalars
 at the Tevatron   \\ [3mm]}
\vskip 1.cm
{\large A.G.~Akeroyd\footnote{A.G.Akeroyd@durham.ac.uk} }
\vskip .4cm
{\it Department of Mathematical Sciences, Centre for Particle
Theory,\\ University of Durham, \\
Durham DH1 3LE, England. }\\
\vskip 1cm
\end{center}

\begin{abstract}
Charged Higgs scalars light enough to contribute to top quark decays are
possible in various non--minimal Higgs models. We show that such a decay
would be consistent with the current Tevatron data, and will remain
hidden until a larger luminosity can be achieved.
 \end{abstract}
\vfill

\end{titlepage}
\newpage
\section{Introduction}
The CDF \cite{Abe} and D0 \cite{Abac} Collaborations have recently announced
 strong evidence for the existence of the top
quark, the isospin partner to the $b$ quark required in the
Standard Model (SM), using
$67 \;{\rm pb}^{-1}$ and $50 \;{\rm pb}^{-1}$ data samples respectively
 of $p\overline p$ collisions
at $\sqrt s=1.8$ TeV. A signal consistent with $t\overline t \to
W^+W^-b\overline b$ has been observed, exceeding the background
prediction by $4.8\sigma$ \cite{Abe} and $4.6\sigma$ \cite{Abac}.
 In Refs. \cite{Abe}, \cite{Abac} the branching ratio BR ($t\to
 Wb$) is taken to be $100\%$, and this is a valid assumption in the minimal
SM (i.e. one Higgs doublet); the decays $t\to cZ$, $t\to uZ$ are
absent at tree level due to the GIM mechanism, and the charged current
processes $t\to Ws$, $t\to Wd$ are
negligible due to heavy CKM matrix suppression ($|V_{ts}|\approx
|V_{td}|\approx 0$).

However, if one enlarges the Higgs sector (i.e. non--minimal
SM\footnote{Defined by assuming no new particles apart from Higgs bosons.})
by adding more doublets and/or triplets, then charged Higgs bosons are
predicted. Such structures are required in many extensions of the SM
(e.g. SUSY, left--right symmetric models), but can also be considered
purely in the context of the non--minimal SM. The vertex $Htb$
is usually predicted in such models and could be quite large due to Higgs
bosons coupling
in proportion to mass. In some extended models (although not all),
$M_H\le m_t-m_b$ is still allowed by current electroweak precision
tests, and if such a light $H^{\pm}$ exists then on--shell $t\to Hb$
decays will occur. This would provide an alternative decay channel for the
top quark and is an option not considered in Refs. \cite{Abe},\cite{Abac}.
It is the
aim of this work to examine whether the presence of a light $H^{\pm}$ is
compatible or not with the CDF data. We shall consider in particular the
case of $M_H\le 80$ GeV i.e. $H^{\pm}$ within the discovery range
of LEP2.

The paper is organised as follows. In Section 2 we briefly
review the various extended Higgs models that may contain a light $H^{\pm}$.
Section 3 examines how significant the channel $t\to Hb$ can be,
while Section 4 studies how one would search for
 $H^{\pm}$. In Section 5 we apply the analyses of Sections 3
and 4 to the current data sample from the Tevatron, while Section 6
considers prospects at an upgraded Tevatron and at the Large Hadron
Collider (LHC). Finally Section 7 contains our conclusions.

\section{Extended Higgs Sectors}

The minimal SM consists of one Higgs doublet ($T=1/2$, $Y=1$), although
extended sectors can be considered and have received substantial
attention in the literature. For a general review see Ref. \cite{Gun}. There
are two main constraints on such models:
\begin{itemize}
\item[{(i)}] There must be an absence of flavour changing neutral
currents (FCNC).
\item[{(ii)}] The rho parameter, $\rho=M^2_W/(M^2_Z\cos^2\theta_W)$, must
be very close to one.
\end{itemize}
Condition (i) is satisfied by constraining the Yukawa couplings to the
fermions \cite{Wein}. Condition (ii) requires models with only doublets, to
which any number of  singlets ($T=0$, $Y=0$) can be added. Models with
triplets ($T=1$) can also
be considered, although obtaining $\rho\approx 1$ is achieved in a less
natural way than for cases with only doublets.

The theoretical structure of the two--Higgs--doublet model (2HDM) is well
known \cite{Gun}, while the general multi--Higgs--doublet model (MHDM)
\cite{Gross}
has received substantially less attention. In the MHDM it is conventional
 to assume that one of the charged scalars is much lighter than
the others and thus dominates the low--energy phenomenology\footnote{In a
model with $N$ doublets there exists $(N-1)$ $H^{\pm}$s.}.  The relevant
part of the Lagrangian for the 2HDM and MHDM can be written as \cite{Gross}:
\begin{equation}
{\cal L} =(2\sqrt2G_F)(X\overline{U}_LVM_DD_R+Y\overline{U}_RVM_UD_L
+Z\overline{N}_LM_EE_R)H^++h.c.
\end{equation}
Here $U_L$, $U_R$ ($D_L$, $D_R$) denote left-- and right--handed up (down)
type quark fields, $N_L$ is the left--handed neutrino field,
and $E_R$ the right--handed charged lepton field. $M_D$, $M_U$, $M_E$ are the
diagonal mass matrices of the down type quarks, up type quarks and
charged leptons respectively. $V$ is the CKM matrix, and  $X$, $Y$ and
$Z$ are coupling constants (see below).

The CP conserving 2HDM which is usually considered in the literature
\cite{Gun} contains an important parameter
\begin{equation}
\tan\beta=v_2/v_1
\end{equation}
with $v_1$ and $v_2$ being real
vacuum expectation values (VEVs) of the two Higgs doublets and $0\le
\beta\le \pi/2$.
There are 4 variants of the 2HDM depending on how the doublets are coupled
to the fermions. Their coupling constants are given in Table 1 \cite{Bar}.
\begin{table}[htb]
\centering
\begin{tabular} {|c|c|c|c|c|} \hline
 & Model I & Model I$'$ & Model II & Model II$'$  \\ \hline
$X$ & $-\cot\beta$ & $-\cot\beta$ & $\tan\beta$ & $\tan\beta$ \\ \hline
$Y$ & $\cot\beta$  & $\cot\beta$ & $\cot\beta$ & $\cot\beta$ \\ \hline
$Z$ &  $-\cot\beta$  & $\tan\beta$  & $\tan\beta$  & $-\cot\beta$ \\ \hline
\end{tabular}
\caption{The values of $X$, $Y$ and $Z$ in the 2HDM.}
\end{table}
In the MHDM $X$, $Y$ and $Z$ are {\it
arbitrary} complex numbers.
It follows that combinations of  parameters like $XY^*$
 have different values
 depending on the model under consideration, thus leading to
phenomenological differences. This has important consequences,
particularly when one calculates loop diagrams involving $H^{\pm}$. One
such decay, $b\to s\gamma$, is sensitive to charged scalars and has
recently been observed for the first time by the CLEO Collaboration. The
value for the branching ratio was measured to be \cite{CLEO}
\begin{equation}
{\rm BR} (b\to s\gamma)= (2.32\pm 0.57\pm 0.35)\times 10^{-4}
\end{equation}
which corresponds to \begin{equation}  1\times 10^{-4}\le {\rm
BR} (b\to s\gamma)\le 4.2\times 10^{-4} \; \; (95\%\; cl) \; .\end{equation}
The theoretical calculation of the branching ratio appears in Refs.
\cite{Kraw}, \cite{Hou}, \cite{Rizz}, \cite{Grin}.
 From this it can be shown \cite{Gross}, \cite{Hew}, \cite{AkeStir} that the
above bound
constrains $M_H$ from the 2HDM (Model II and II$'$) to be $\ge 260$ GeV,
while no bound can be obtained on $M_H$ from the 2HDM (Model I
and I$'$) and the MHDM. Hence it is possible that an on--shell $H^{\pm}$ from
these latter models contributes to top decay, and may even be light
enough to be detectable at LEP2 \cite{AkeStir}, \cite{Sop}. Also, we note
that
the most popular model with Higgs isospin triplets (HTM) \cite{Geo},
\cite{Wud}, predicts a charged scalar
$H^{\pm}_3$ which has exactly the same couplings as $H^{\pm}$ (2HDM,
Model I), and thus may also contribute to top decay and/or be detectable
at LEP2 \cite{Ake}. For all these $H^{\pm}$s there exists a lower bound from
LEP \cite{Rev} of  $M_H\ge 41.7$ GeV.

An important constraint on $\cot\beta$ is obtained from precision
measurements of the $Z\to b\overline b$ vertex. Charged scalars with a tree
level $Htb$ coupling contribute to this decay, and Ref. \cite{Gross} shows
that \begin{equation} |\cot\beta|\le 0.8.\end{equation} This bound is for
$m_t=180$ GeV and $M_H\le 200$ GeV, and provides a stronger constraint
on $|\cot\beta|$ than can be obtained from the decay $b\to s\gamma$.

In order to search for $H^{\pm}$ one needs to consider its decays.
The various branching ratios (BRs) are shown in Table 2, with $l$
referring to either an electron or a muon.\begin{table}[htb]
\centering
\begin{tabular} {|l|c|c|c|} \hline
 & Jets & $\tau\nu_{\tau}$ & $l\nu_{l}$  \\ \hline
 $H^{\pm}$ (2HDM, Model I) & $67\%$  & $33\%$ & $0\%$  \\ \hline
 $H^{\pm}$ (2HDM, Model I$'$) & $\le 46\%$  & $\ge 54\%$  & $0\%$ \\ \hline
$H^{\pm}$ (MHDM)& $0\to 100\%$ & $0\to 100\%$ &  $0\%$ \\ \hline
$W^{\pm}$ & $67\%$ & $11\%$ & $22\%$ \\ \hline
\end{tabular}
\caption{The branching ratios for $H^{\pm}$ and $W^{\pm}$.}
\end{table}
For the 2HDM (Model I) the BRs are independent of
$\tan\beta$, while in the 2HDM (Model I$'$) the constraint of Eq. (5) creates
the inequalities
in Table 2. In the MHDM the BRs are dependent on the
arbitrary parameters $X$, $Y$ and $Z$, and thus span the full range $0\to
100\%$.

\section{Top Quark Decay to Charged Higgs}
In this section we evaluate how competitive the decay channel $t\to
Hb$ can be compared to the conventional $t\to Wb$. The Feynman rule for the
$Htb$ vertex is given by \cite{Gun} \begin{equation}
{igV_{tb}\over 2\sqrt2M_W}[m_bX(1+{\gamma}_5)+m_tY(1-{\gamma}_5)]\;.
\end{equation}
Here $V_{tb}$ is a CKM matrix element and $g$ is the usual SU(2) gauge
coupling. The 2HDM (Models I and I$'$) requires the replacements
$X=-\cot\beta$ and $Y=\cot\beta$.
 From Eq. (6) one can evaluate the partial width and thus BR $(t\to H^+b)$
\cite{Gun}: \begin{equation}
{{\rm BR}(t\to H^+b)\over {\rm BR}(t\to W^+b)} = {p_{H^+}\over p_{W^+}}
\times
{\left[(m^2_b+m^2_t-M^2_H)(m^2_b+m^2_t)-4m^2_bm^2_t\right]\cot^2\beta
\over M^2_W(m^2_t+m^2_b-2M^2_W)+(m^2_t-m^2_b)^2}\;.
\end{equation}
Here $p_{H^+}$ and $p_{W^+}$ refer to the magnitude of the three momentum
of the $H^+$ and $W^+$ measured in the rest frame of the top quark. We take
$m_t=174$ GeV, $m_b=5$ GeV and $M_W=80.3$ GeV.

\begin{figure}[htbp]
\begin{center}
\mbox{\epsfig{file=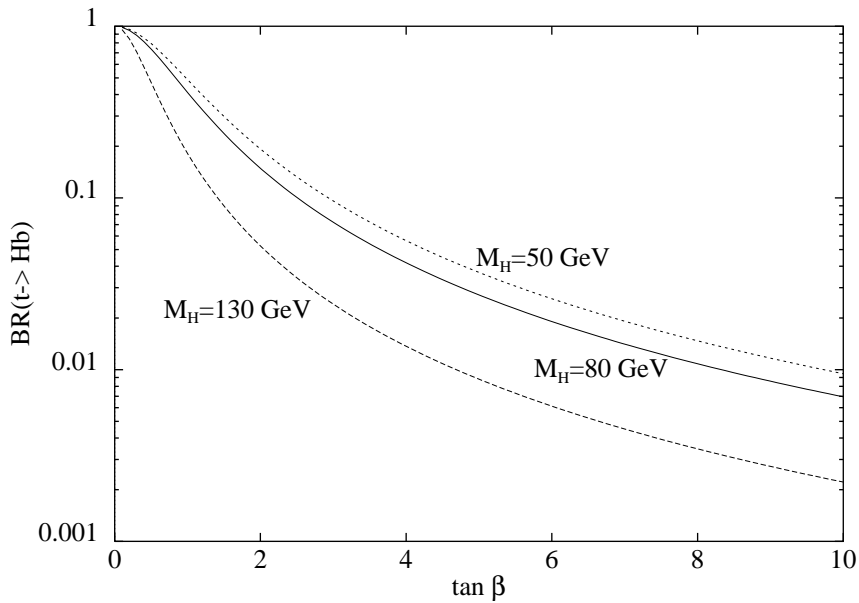,angle=-90,height=9cm}}
\end{center}
\vspace{-5mm}
\caption{BR $(t\to Hb)$ as a function of $\tan\beta$ for $M_H=50$,
$80$ and $130$
GeV.}
\label{Fig:Top6}
\end{figure}

In Figure \ref{Fig:Top6} we plot BR $(t\to H^+b)$ as a function of
$\tan\beta$ for $M_H=50$
GeV (comfortably in the LEP2 range), for
$M_H=M_{W}$ (at the edge of the LEP2 range), and for $M_H=130$ GeV (out
of LEP2 range). We recall that Eq.
(5) states $|\cot\beta|\le 0.8$ or equivalently $|\tan\beta| \ge 1.25$.
Using this bound we see from Figure 1 that large values of BR
$(t\to H^+b)$ are still possible e.g. for $M_H=50$ GeV and
$\tan\beta=1.25$, BR $(t\to H^+b)= 38\%$. As $\tan\beta$ increases  BR
$(t\to H^+b)$ falls, dropping to $\approx 1\%$ at $\tan\beta=10$.
Hence for large enough $\tan\beta$ and/or $M_H$ the top decay to
 charged Higgs will be
difficult to detect experimentally. We shall be particularly interested
in the case of $M_H\le M_W$ i.e. for $H^{\pm}$ in
the discovery range of LEP2.

In the MHDM, BR $(t\to H^+b$) depends not on $\tan\beta$ but instead on
$|X|$ and $|Y|$. However, it is easily verified from Eq. (7) (unless
$|X|$ is unnaturally large -- see below) that
the strongly dominant part is proportional to $|Y|^2$ which has the
same
constraint as $\cot\beta$ (Eq. 5). Therefore Figure 1 can be used for
the MHDM, with the replacement $\tan\beta\to |Y|^{-1}$ on the ordinate axis.
We note that there exists an unnatural region of parameter space, $|X|\ge 30$
and $|Y|\le
0.1$,\footnote{There is an experimental constraint on the product $|X||Y|$
coming
from the measurement of $b\to s\gamma$. Ref. \cite{Gross} uses $|X||Y|\le 4$
for $M_H\le 80$ GeV, although with the new measurement in Eq. (3),
$|X||Y|\le 3.3$ is more accurate.}
which would allow larger BRs in the MHDM for a given $M_H$ than is ever
possible in the 2HDM. We shall not consider this possibility.

\section{Signature of the Charged Higgs}
We shall consider two ways with which to detect $H^{\pm}$ from
top decays. The first method is to search in the lepton channel. We
see from Table 2 that the decays of $H^{\pm}$ violate lepton universality
due to a preference to couple to the heaviest lepton ($\tau$). This is in
contrast to the lepton decays of $W^{\pm}$. The second method is to
 search for the quark decays of $H^{\pm}$ by reconstructing the
invariant masses of the jets. A significant peak separate from that of
$W^{\pm}$ would be distinctive. A search using the di--lepton channel
$t\overline t\to H^+H^-b\overline b\to llX$ (via $H^{\pm}\to
\tau\nu_{\tau}\to l\nu_l\overline \nu_{\tau}\nu_{\tau}$) has been
performed \cite{Abe3}    although we shall not consider this method here.
We shall focus first on the $\tau\nu_{\tau}$ decays of $H^{\pm}$. A
previous search in this channel has been performed at the Tevatron using
a $4\; {\rm pb}^{-1}$ data sample \cite{Abe2}. Here BR ($t\to Hb$)=100\%
was
assumed which we know now to be untrue since $m_t>M_W+m_b$, and so on--shell
$t\to Wb$ decays are definitely present.  In Ref. \cite{Abe2} the process
$t\overline t\to H^+H^-b\overline b\to
\tau^+\tau^-\nu_{\tau}\overline \nu_{\tau}b\overline b$ was searched for
with the
trigger being missing $E_T\ge 25$ GeV. From Figure 1 we see that BR ($t\to
H^+b)$ can be as low as $1\%$, and in such cases the
above detection method is certainly not relevant for future searches. One
must rely on the process
$t\overline t\to H^{\pm}W^{\mp}b\overline b\to
l\nu_l\tau\nu_{\tau}b\overline b$, with a
hard isolated lepton ($l=e$ or $\mu$) to act as a trigger \cite{Abe}, \cite
{Abac}. We
also require a $\tau$ jet to which various cuts will be made
\cite{Abe2}. Non
$t\overline t$ backgrounds, although sizeable, are substantially reduced
when the above requirements are applied. The main
background is from $t\overline t\to W^+W^-b\overline b\to
l\nu_l\tau\nu_{\tau}b\overline b$ which
fakes the signal from $H^{\pm}W^{\mp}$ and is usually the more common top
decay
process. Fortunately this background can be reduced by making use of the
polarisation of the $\tau^{\pm}$. To be specific we shall focus on $W^-$ and
$H^-$, thus following the notation of Ref. \cite{Bull}.

Weak vector bosons couple to left--handed fermions and right--handed
antifermions, and thus $W^-\to
\tau^-_L\overline \nu_{\tau}$. However this is not so for the charged Higgs
($H^-$) which
couples to right handed fermions via $H^-\to \tau^-_R\overline \nu_{\tau}$.
This is a consequence of
the Yukawa couplings. It has been known for some time that the momentum
distribution of the decay products from $\tau^{\pm}$ depend on its
polarisation \cite{Tsai}. Ref. \cite{Bull} analyses these differences for
the hadronic
decays $\tau^-\to \pi^-\nu_{\tau}\;, \rho^-\nu_{\tau}\;,
a_1^-\nu_{\tau}$.
It concludes that the energetic particles coming from $W^-$ are
primarily $a^-_{1T}$ and $\rho^-_T$, with the subscript $T$ referring to a
transversely polarised meson. In comparison, the energetic particles
coming from $H^-$ are $\rho^-_L$, $a^-_{1L}$, and $\pi^-$ with $L$
referring to a longitudinally polarised meson. The case of the energetic
$\pi^-$ from $H^-$ can easily be understood in the following way. Due to
polarisation of the parent $\tau^-$, spin conservation favours emission
of the $\pi^-$ in the direction of motion of the  $\tau^-$. This results
in the $p_T$ spectrum in the context of $t\overline t$ production of the
$\pi^-$ being boosted compared to those
pions coming from $W^-$ decay (in the latter case the $\pi^-$ prefers to be
emitted {\it opposite} to the direction of the parent $\tau^-$).
Therefore the fragmentation function of $\pi^-$ from $\tau^-_R$ is
harder than that from  $\tau^-_L$, and so a hard pion is a signal of
$H^-$ decay.

The subsequent decays of $\rho^-\to \pi^0\pi^-$ and $a^-_{1} \to
\pi^-\pi^-\pi^+$, $ \pi^0 \pi^0\pi^-$ have different energy distributions
depending on the polarisation of the parent meson. Ref. \cite{Bull} shows
that
for the $\rho^-_L$ the final state pions will be much more asymmetric in
energy than those originating from $\rho^-_T$. For the case of the
$a^-_{1T}$ the three pions tend to share the energy equally, while for the
$a^-_{1L}$ one or two of them are soft with the rest hard. Hence the
signature of the $H^-$ is a final state of one or more energetic pions.

It is straightforward to evaluate the theoretical prediction for the
number of energetic pions from top decay. We take first the case of
$t\overline t \to W^+W^-b\overline b$. One of the $W$s is decayed to
$l\nu_{l}$ to act as a trigger for the event, while the other is decayed
to $\tau\nu_{\tau}$. The theoretical prediction is: \begin{equation}
N^{WW}_{l\tau}=2(1-B_H)^2N_{t\overline t}\times {\rm BR}\;(W\to
l\nu_l)\times {\rm BR} \;(W\to \tau\nu_{\tau})e_{l\tau}\;.
\end{equation}
Here $B_H$ is a shorthand for BR ($t\to Hb$) and $N_{t\overline t}$ is
the number of $t\overline t$ pairs produced. The parameter $e_{l\tau}$
is an overall
efficiency for detecting an energetic pion from the above event,
incorporating any kinematical cuts. It can be written as \begin{equation}
e_{l\tau}=  e_l\times \sum _{i=1}^3 e_{i,\,\pi}\times {\rm
BR}\;(\tau^-\to \pi^- \nu_{\tau}+X_i) \;. \end{equation}
The summation is over the three possible decays of $\tau^-$ to pions: $X_2$,
$X_3$ are extra pions originating from  $\tau^-\to \rho^-\nu_{\tau}$,
 and $a_1^-\nu_{\tau}$ respectively, while $X_1\equiv 0$. For the case of
$t\overline t
\to W^{\pm}H^{\mp}b\overline b$ theory predicts \begin{equation}
N^{WH}_{l\tau}=2(1-B_H)B_HN_{t\overline t}\times {\rm BR}\;(W\to
l\nu_l)\times {\rm BR} \;(H\to \tau\nu_{\tau})e_{l\tau}'\;,
\end{equation}
with $e_{l\tau}'$ being different to $e_{l\tau}$. This latter difference
arises due to the polarisation of the parent $\tau$ and thus
$e_{i,\,\pi}\ne
e_{i,\,\pi}'$. There are two ways of searching for violation of lepton
universality. The
first method (applied in Ref. \cite{Hubb} to the SSC) requires an isolation
cut on the energetic pion. This eliminates virtually all of the QCD
background, although also has the disadvantage of removing most of the hard
pions originating from $\rho^-$ and $a_1^-$.  A second method (suggested
in Ref. \cite{Roy}) aims to include these latter decays by rejecting the
isolation cut, and instead making a cut on $\Delta E_T\equiv
|E_T^{\pm}-E_T^0|$ i.e. the total $E_T$ carried by the charged pions minus
the total $E_T$ carried by the neutral pions. The signal favours large
$\Delta E_T$. This
method keeps more of the signal than would be kept by making the
isolation cut, although
the QCD background is larger. The lepton trigger efficiency, $e_l$
and $e_l'$,
depends only on $m_t$ and is therefore the same for both events.
We will ignore the case of both top quarks decaying to a charged
Higgs ($N^{HH}_{l\tau}$) for the following reasons:
\begin{itemize}
\item[{(i)}] BR $(t\to Hb)$ could well be
very small (see Figure 1).
\item[{(ii)}]  The trigger here
would be $H\to \tau\nu_{\tau}\to
l\nu_l\nu_{\tau}\overline\nu_{\tau}$. This process would  be suppressed in
models with lower BR
($H\to \tau\nu_{\tau}$) i.e. 2HDM (Model I), HTM and various MHDM, coupled
with the fact that BR $(\tau\to l\nu_l\nu_{\tau}) \approx 36\%$.
 \item[{(iii)}] The trigger decay in (ii) predicts $l$
with less $p_T$ than for the $l$ originating from $W\to l\nu_{l}$. Hence
it is less likely to pass the lepton trigger cut.\end{itemize}
Reasons (ii) and (iii) also allow us to ignore events $N_{\tau\tau}^{WW}$
which fake $N_{l\tau}^{WW}$.
We can quantify the amount of deviation from universality that would be
caused by a non--zero $N^{WH}_{l\tau}$:
\begin{equation}
N_{\sigma}={N^{WH}_{l\tau}\over \sqrt{N^{WH}_{l\tau}+N^{WW}_{l\tau}}}\;,
\end{equation}
with $N_{\sigma}$ being the number of standard deviations by which the
observed number of hard, isolated pions exceeds that predicted from
universality.
This concludes our account of the search for $H\to \tau\nu_{\tau}$.

The second method is to consider the $H\to cs$ decay and
reconstruct the invariant masses of the jets. This method would be needed
for the case of a `leptophobic' Higgs i.e. BR $(H\to \tau\nu_{\tau})\to
0\%$ and
BR $(H\to cs)\to 100\%$, which is possible in the MHDM for large $|Y|$
and small $|X|$, $|Z|$. The trigger will again be a
hard isolated lepton originating from $W\to l\nu_l$. Three jets need to
be reconstructed in the opposite hemisphere to the lepton (i.e. the jets
originating from $t\to Hb\to csb$), with one of them a tagged $b$ jet
(efficiency $e_b'$).
Ref. \cite{Hubb} deals with this method applied to the SSC, and the analysis
is also relevant for the Tevatron with a few minor modifications e.g. the
$p_T$ cuts
given in Ref. \cite{Hubb} are larger than those required at the Tevatron
due to
$\sqrt s$ at the SSC being much greater. Adding the invariant masses of the
 non $b$ jets will result in a peak centered on
$M_H$. However, one must be careful here because in the MHDM it is
possible to have a large BR $(H\to cb)$ \cite{AkeStir}; in this
scenario there is a chance of the wrong $b$ jet being tagged and therefore
the wrong
jets would be used to reconstruct the invariant mass of $H^{\pm}$.
However, BR $(H\to cb)\ge 30\%$ would be needed to cause significant
problems here, and  Ref. \cite{AkeStir} shows that the parameter space
for this large branching ratio is quite small.

Requiring that the three--jet invariant mass is centered on $m_t$ reduces
the background further. A clear peak would be strong evidence for
$H^{\pm}$. Ref. \cite{Hubb} suggests the following simple rule for this
method to provide a significant peak at $M_H$:
\begin{equation}
{\rm BR}\;(t\to Hb)\times{\rm BR}\;(H\to cs)\ge 0.05\,.
\end{equation}
This result was obtained by using the ISAJET Monte Carlo, comparing the
size of the $H^{\pm}$ peak to the $W^{\pm}$ background for different
values of $\tan\beta$ and $M_H$. It is only valid for a large
event sample (at the SSC $N_{t\overline t}= 10^7\to 10^8$), and so will
be more relevant at an upgraded Tevatron and/or LHC. If Eq. (12) is
satisfied,
 the statistical significance of the peak at $M_H$ is greater with a
larger event sample, i.e. at the LHC. We note that Eq. (12) was derived
using SSC detection efficiencies; however it will still be valid at the
Tevatron to a good approximation.\footnote{The
ratio $e_b/e_b'$ contributes to the relative size of the peaks at $M_W$
and $M_H$. This is the only factor which may invalidate Eq. (12) for use
at the Tevatron.}

In this section we have studied two direct ways with
which one may search for $H^{\pm}$. In the next section we shall apply
these techniques to the current data sample from the Tevatron to see if
a light ($\le 80$ GeV) charged Higgs boson could have been observed.

\section{Analysis at the Tevatron}
The current CDF data sample at the Tevatron [1]
contains $67\;{\rm pb}^{-1}$. For $\sigma_{t\overline t}$
we shall use the MRS(A) partons \cite{Stir1} which suggests a value of
4.5 pb if $m_t=174$ GeV\footnote{ We note that a recent calculation
\cite{Berg}
suggests  $\sigma_{t\overline t}=5.52^{+0.07}_{-0.45}$ pb.} \cite{Stir2}.
Ref. \cite{Ell} estimates
that the theoretical error in  $\sigma_{t\overline t}$ is $\pm 30\%$.
We can now evaluate the expected number of $t\overline t$ pairs, and then
use Eqs. ($8\to 10$) to predict the number of hard pions for
various values of BR $(t\to Hb)$.

We shall take $N_{t\overline t}= 300$; from Table 2 one has BR $(W\to
l\nu_l)=22\%$ and BR $(W\to \tau\nu_{\tau})=11\%$. If there is no charged
Higgs boson i.e. BR $(t\to Wb)=100\%$ then one has from Eq. (8):
\begin{equation}
N^{WW}_{l\tau}=14.5\times e_{l\tau}\;.
\end{equation}
The efficiency $e_{l\tau}$ is given by Eq. (9), and we shall discuss its
value below.
The prediction for $N^{WH}_{l\tau}$ is given by Eq. (10). We shall take BR
$(t\to Hb)=38\%$ for illustrative purposes i.e. the maximum BR for $M_H=50$ GeV
(see Figure 1). Thus we obtain:
\begin{equation} N^{WH}_{l\tau}=31.1\times {\rm BR}\; (H\to \tau\nu_{\tau})
\times e_{l\tau}'\;.
\end{equation}
We shall consider the search strategy using an isolation cut on the
charged pion. As mentioned before this eliminates most of the QCD
background.
The charged pion is required to contain at least $80\%$ of the
energy in a cone of size $\Delta R=0.2$, as well as possessing $E_T\ge 20$
GeV. This removes most of the
$W$ background contribution from $\rho$ and $a_1$ and most of
the signal contribution from  $a_1$  \cite{Roy}. Therefore we may use the
following approximations in Eq. (9):
\begin{equation}
e_{2,\,\pi}=e_{3,\,\pi}=e_{3,\,\pi}'=0.
\end{equation}
The efficiency $e_{l\tau}'$ will be larger
than $e_{l\tau}$ for the following reasons. Most importantly,
$e_{1,\,\pi}'> e_{1,\,\pi}$
i.e. pions originating from $H^{\pm}$ are more energetic (see Section 4) and
are more likely to pass an energy cut than those from $W^{\pm}$. Using
the figures in Ref. \cite{Roy} we can approximate $e_{1\,,\pi}=0.26$.
Also we want to maximise the signal for illustrative purposes and so take
\begin{equation} e_{1\,,\pi}'=1.0\;,\; e_{2,\,\pi}'=1.0. \end{equation}
The $E_T>20$ GeV cut has little effect on $
e_{1\,,\pi}'$ and $ e_{2,\,\pi}'$; the isolation cut hardly affects
the $\pi$ contribution but removes part of the
$\rho$ contribution. Hence in
reality  $ e_{2,\,\pi}'<1$. For heavier $H^{\pm}$ the pion
is on average more energetic and so $e_{i,\,\pi}'$ will increase. The
trigger efficiencies for an isolated, $p_T\ge 20$ GeV lepton ($e_l$, $e_l'$)
are close to 100\% \cite{Roy1}. Combining
the above efficiencies, it is clear that $e_{l\tau}'>e_{l\tau}$.
 From Table 2 we can obtain the value of BR $( H\to \tau\nu_{\tau})$
for the various Higgs models. The maximum signal is obtained when BR $(H\to
\tau\nu_{\tau})\to
100\%$\footnote{This situation is only possible in the
 2HDM (Model I$'$) for $\tan\beta\ge 3$ and in the MHDM. However, in the
calculation we also use
 BR $(t\to Hb)=38\%$, and both these BRs can
only be satisfied simultaneously in the MHDM for
$|X|^{-1}=-|Y|^{-1}=\tan\beta=1.25$ and $|Z|$ large.}
and we will use this in our calculation. The remaining parameter values
needed are BR ($\tau\to \pi\nu_{\tau})=12.5\%$ and BR ($\tau\to
\rho\nu_{\tau})=24.0\%$.

Using the above values for
efficiencies and BRs in Eqs. (13) and (14) we find
\begin{equation}
N^{WW}_{l\tau}=0.48\;\; {\rm and}\;\; N^{WH}_{l\tau}=11.35.
\end{equation}
Using Eq. (11) we see that to obtain a $5\sigma$ signal $
N^{WH}_{l\tau}=25$ is required and this is far from the prediction of 11.35.
Also, we used the most favourable values for BR $( H\to
\tau\nu_{\tau})$, BR $(t\to Hb)$ and $e_{l\tau}'$. Lower values for these
parameters would decrease $N^{WH}_{l\tau}$ further e.g. for $M_H=50$ GeV,
$\tan\beta=3$ and BR $(H\to \tau\nu_{\tau})=33\%$ (Model I) we find
$N^{WH}_{l\tau}=1.45$. Therefore we conclude that, over the vast majority
of parameter ($M_H$, $\tan\beta$) space,
$H^{\pm}$ would not provide a statistically significant signal given the
current data at the Tevatron. The other method available to us exploits the
quark decays of $H^{\pm}$. From Eq. (10), with BR $(H\to cs)$ replacing
BR $(H\to \tau\nu_{\tau}$), we again see that too few events are produced.

The conclusion of the above analyses is that a direct signature of $H^{\pm}$
cannot be obtained using
the current data sample from the Tevatron. However an indirect signature
might be possible. The top quark search [1] relied on an excess of $W +
4$ jet events over the QCD background. The events searched for
consisted
of one $W$ providing the hard lepton trigger, while the other decays to
two quarks.  A light $H^{\pm}$ can mimic this signal
although it would suppress the expected number of events in this channel
($N_{\rm W4j})$, and
so increase the experimentally measured cross section for $t\overline t$
production. We shall now quantify this for various values of BR
($t\to Wb$). The experimentally measured cross--section ($\sigma_{\rm
exp}$) can be written as
\begin{equation}
\sigma_{\rm exp}={N_{W4j}\over {ke}}
\end{equation}
with $e$ ($=e_le_b$) being a detection efficiency for the $W$ plus 4 jets
channel,
and $k$ is a parameter depending on branching ratios and machine luminosity.
To a good approximation the efficiency $e$ remains the same whether we
input a charged Higgs
 or not.\footnote{Ref. \cite{Hubb} shows that $e_l$ is
independent of $M_H$ while $e_b$ depends relatively weakly on $M_H$,
especially in the mass range 50 GeV $\le M_H \le$ 80 GeV which
gives larger BR ($t\to Hb$).} However, the presence of $H^{\pm}$ will
change
$k$ while $N_{W4j}$ is a fixed experimental measurement irrespective of
which theory we are considering. In the following analysis we shall
consider Model I which has a larger BR $(H^{\pm}\to$ jets) than Model
I$'$ (see Table 2). Therefore the latter model would suppress $N_{W4j}$
more. In the MHDM BR ($H^{\pm}\to$ jets) may take any value between
$0\%$ and $100\%$. Table 3 shows how $k$ (for
$67\; {\rm pb^{-1}}$) varies with different values of BR ($t\to Wb$).
\begin{table}[htb]
\centering
\begin{tabular}  {|c|c|} \hline
BR $(t\to Wb$) & $k\;({\rm pb}^{-1})$   \\ \hline
100\% & 19.75  \\ \hline
99\% & 19.55  \\ \hline
95\% & 18.76 \\ \hline
90\% & 17.76 \\ \hline
62\% & 12.24  \\ \hline
\end{tabular}
\caption{The variation of $k$ with BR $(t\to Wb$).}
\end{table}
Ref. [1] assumes BR $(t\to Wb)=100\%$ and $\sigma_{\rm exp}$ is
measured to be $6.8^{+3.6}_{-2.4}$ pb. From Eq. (18) one can see that
if  BR $(t\to Wb)= x\%$ with $x<100$ the CDF measurement would be
 scaled by the ratio
$k(100\%)/k(x\%)$.
The lower error bar on $\sigma_{\rm exp}$ currently lies within
the region of $\sigma_{\rm theo}$ $(=4.5\pm 1.35$ pb with $m_t=174$ GeV),
and the inclusion of
 $t\to Hb$ decays will shift the $\sigma_{\rm exp}$ data point to higher
values. If BR ($t\to Hb)=24.8\%$ then the
lower error bar on $\sigma_{\rm exp}$ lies just outside the maximum value
of $\sigma_{\rm theo}$ (5.85 pb). Therefore lower
branching
ratios than this are more desirable. From Figure 1 we see that a
considerable parameter
space exists for BR ($t\to Hb)\le 5\%$; in such cases $\sigma_{\rm exp}$
would only be increased by $\approx 5\%$
and so a sizeable portion of the lower error bar would still lie within
the region
of $\sigma_{\rm theo}$. Therefore we conclude that top quark decays to
charged Higgs scalars
are consistent with the current CDF data over a very large parameter
space of $\tan\beta$ and $M_H$. Therefore $H^{\pm}$ may lie in the energy
range of LEP2.

The Tevatron will continue to operate until the end of 1995, eventually
collecting 140 ${\rm pb^{-1}}$ of data. From Eqs. (13) and (14) one sees
that this twofold
increase would only provide a statistically significant signal for a
minute part of parameter space. The most favourable choice of parameters
(used previously) gives $N_{\sigma}=4.8$.  Higher luminosity
colliders are required and these are studied in the next section.
  \section{Prospects at an Upgraded Tevatron/LHC}
There is a proposal to increase the luminosity of the Tevatron by an
order of magnitude and it is thought that 2 ${\rm fb}^{-1}$ might be
possible by the year 2000. Further proposals are for future
luminosities of 20 ${\rm fb}^{-1}$ and/or $\sqrt s=4$ TeV. Finally,
by the year 2010 the Large Hadron Collider (LHC) should be operating.
 Table 4 \cite{Yuan} summarises the properties
of these colliders.  \begin{table}[htb]
\centering
\begin{tabular}  {|c|c|c|c|c|} \hline
  & $\sqrt s $ (TeV) &  $\sigma_{t\overline t}$  $({\rm pb})$ &
$L$ $(\rm fb)^{-1}$  & ${\rm Operation}$  \\ \hline
{\rm Tevatron} & 1.8 &4.5 & 2 & 2000 \\ \hline
{\rm Tevatron}*  & 1.8 &4.5 & 20 & $>2000$  \\ \hline
{\rm Di--Tevatron} & 4 &26 & $2\to 20$ &  $>2000$   \\ \hline
{\rm LHC} &  14 & 430 & 100 & 2010 \\ \hline
\end{tabular}
\caption{Important parameters at future colliders.}
\end{table}
We shall now consider the
detection
prospects of $H^{\pm}$ via top quark decay at each of these colliders.
 In
this section we shall concentrate on studying the mass range $80$ GeV $\le
M_H\le m_t-m_b$; this is because LEP2 will search for $M_{H}\le 80$ GeV
\cite{AkeStir}, \cite {Sop} before any of the above colliders will operate.
An analysis of the detection prospects of the minimal SUSY $H^{\pm}$ at the
upgraded Tevatron appears in Ref. \cite{Roy}. The Higgs models we are
studying differ noticeably from the SUSY $H^{\pm}$ with respect to BR
($t\to Hb$). The difference arises because SUSY requires Model II couplings
\cite{Gun}
and thus  BR ($t\to Hb$) has a sinusoidal dependence on $\tan\beta$ with a
minimum at $\tan\beta\approx6$. In contrast we are studying Model I
and I$'$ type couplings which cause BR ($t\to Hb$) to fall with increasing
$\tan\beta$ (Figure
1). Therefore $H^{\pm}$ in these latter models will always be hidden for
large enough $\tan\beta$.

\begin{figure}[p]
\begin{center}
\mbox{\epsfig{file=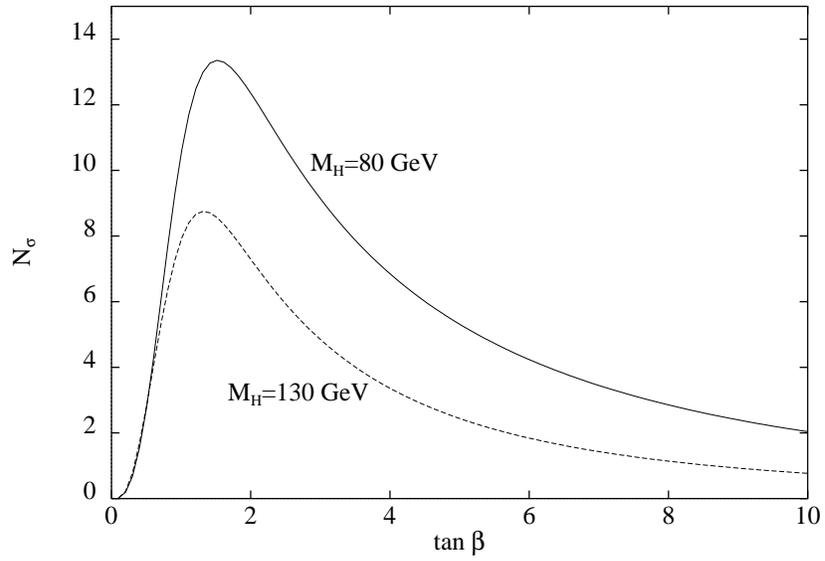,angle=-90,height=8.6cm}}
\end{center}
\vspace{-5mm}
\caption{$N_{\sigma}$ as a function of of $\tan\beta$ for Model I$'$ with
$M_H=80$ and 130 GeV. The Tevatron with $L=2$ fb$^{-1}$ is considered.}
\label{Fig:Top1}
\end{figure}

\begin{figure}[p]
\begin{center}
\mbox{\epsfig{file=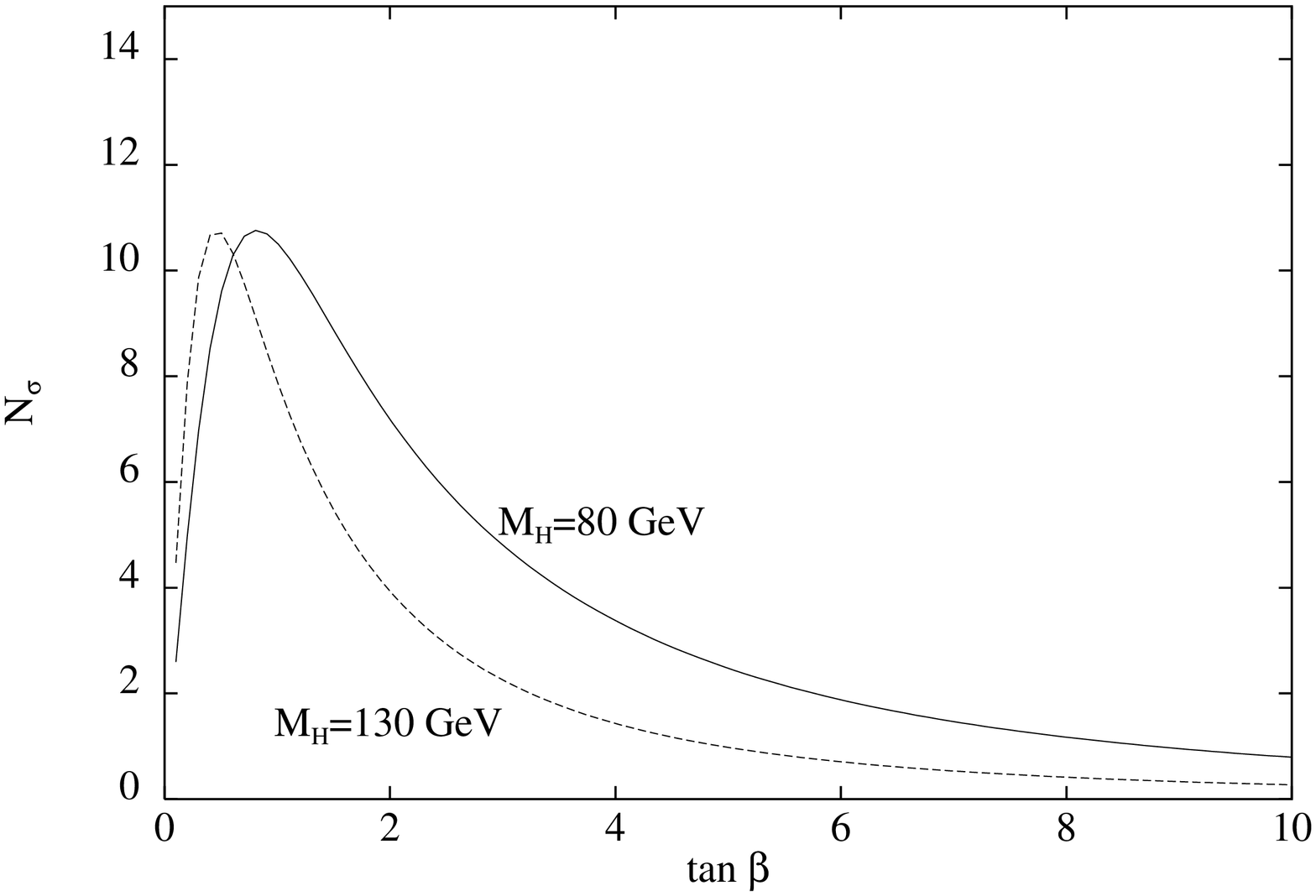,angle=-90,height=8.6cm}}
\end{center}
\vspace{-5mm}
\caption{Same as Figure \protect{\ref{Fig:Top1}} but for Model I.}
\label{Fig:Top}
\end{figure}

Starting with an upgraded Tevatron providing 2 ${\rm fb}^{-1}$, we again
make use of Eqns. (8) and (10) using the same values for
the efficiencies as in Section 5. The number of $t\overline t$ pairs
($N_{t\overline t}$) will now be taken to be $9000$ although there is
still a theoretical error here of $\pm 30\%$. In Figures \ref{Fig:Top1}
and \ref{Fig:Top} we plot
Eq. (11)
which shows the excess (quantified as a number of standard deviations above
that predicted from universality) of high--energy pions
caused by $H^{\pm}$ decay.

Figure \ref{Fig:Top1} (for Model I$'$) shows that for $M_H=80$
GeV, $N_{\sigma}\ge
5\sigma$
is maintained until $\tan\beta\approx 5.5$ while $N_{\sigma}$ $\ge 3\sigma$
is managed until $\tan\beta\approx 8$. For heavier $M_H$ the signal is
obviously weaker; for $M_H=130$ GeV, $N_{\sigma}\ge 5\sigma$ ($\ge
3\sigma$)
is maintained only until $\tan\beta\approx 3 (5)$. For Model I (Figure
\ref{Fig:Top}) the situation is worse due to the inferior
BR $(H\to \tau\nu_{\tau})$.
In the MHDM, BR $(H\to \tau\nu_{\tau})$ can take any value from $0\%\to
100\%$. However the case of BR $(H\to \tau\nu_{\tau})=100\%$
approximates to Figure \ref{Fig:Top1} as $\tan\beta$ increases.\footnote{In
Model I$'$ BR $(H\to \tau\nu_{\tau})\approx 100\%$ for $\tan\beta \ge 3$.}

Prospects are improved at the Tevatron* (20 fb$^{-1})$, Di--Tevatron
(2  fb$^{-1})$  and Di--Tevatron (20 fb$^{-1}$); these would scale
$N_{\sigma}$ by
3.2, 2.4 and 7.6 respectively. The latter collider would maintain a
$5\sigma$ signal in Model I$'$ (Model I) for $M_H=130$ GeV if $\tan\beta\le
11$ (8).

The other detection method available is to reconstruct the
invariant masses of the quark decays of $H^{\pm}$. Certainly this method
will be needed in the case of a leptophobic Higgs (see Section 4) which
is possible in the MHDM.
\begin{figure}[p]
\begin{center}
\mbox{\epsfig{file=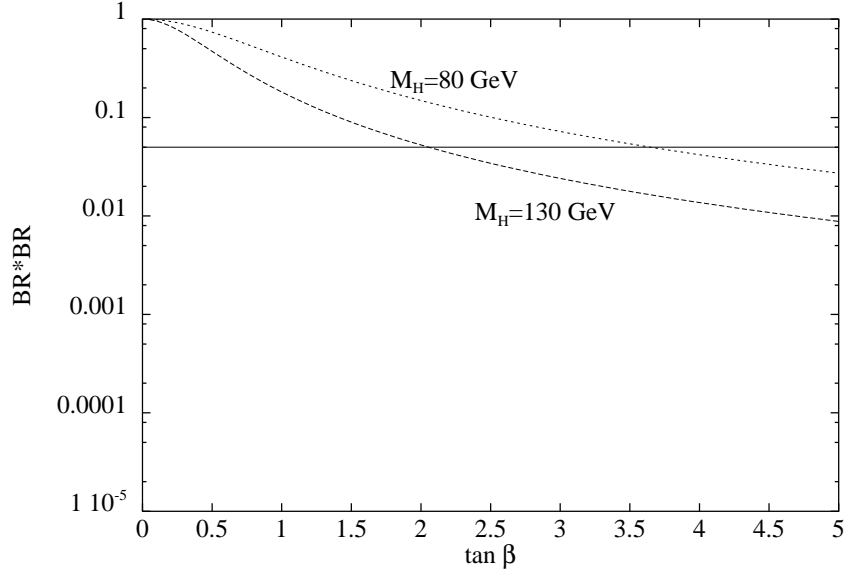,angle=-90,height=8.6cm}}
\end{center}
\vspace{-5mm}
\caption{BR $(t\to Hb)\times$BR $(H\to cs)$ as a function of $\tan\beta$
for a leptophobic $H^{\pm}$ with $M_H=80$ and 130 GeV. The detectable
region lies above the line BR$\times$BR=0.05.}
\label{Fig:Top5}
\end{figure}
\begin{figure}[p]
\begin{center}
\mbox{\epsfig{file=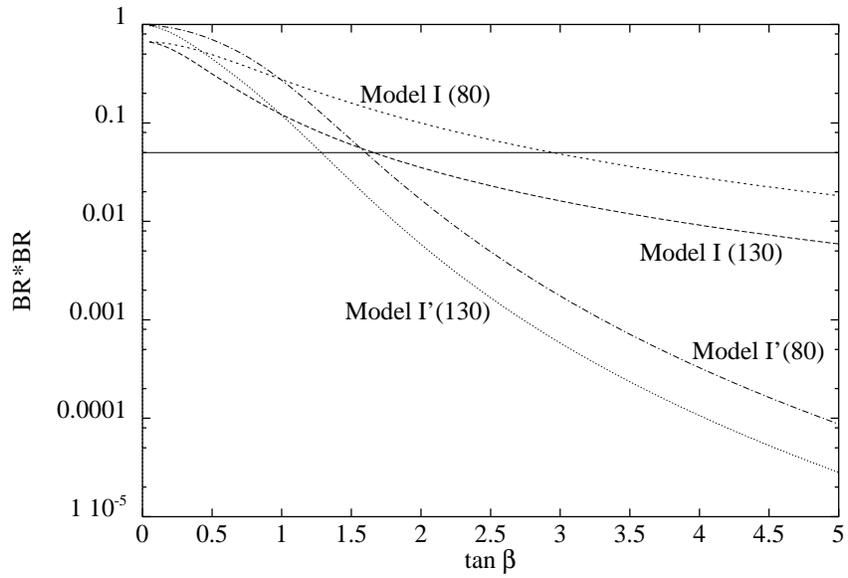,angle=-90,height=8.6cm}}
\end{center}
\vspace{-5mm}
\caption{Same as Figure \protect{\ref{Fig:Top5}} but for Model I and I$'$.}
\label{Fig:Top2}
\end{figure}
Figure \ref {Fig:Top5} shows that for such a Higgs a noticeable peak at $M_H$
can be obtained if $M_H=80$ GeV (130  GeV) for $\tan\beta\le 3.5$ (2).
Figure \ref{Fig:Top2} is the analogous plot for Model I and I$'$;
 it is clear that
in Model I reconstructing the $H\to$ jets channel is competitive
with the previously considered $H\to \tau\nu_{\tau}$ decay (if $L=2$
fb$^{-1}$). A noticeable peak
would be present for $M_H=80$ GeV (130  GeV) if $\tan\beta\le 3$ (1.7).
In comparison from Figure \ref {Fig:Top} we see that the cases of $M_H=80$ GeV,
$\tan\beta =3$ and
 $M_H=130$ GeV,  $\tan\beta=1.7$ both give $N_{\sigma}=5$. Hence the
detection methods are competitive. At the Tevatron* and Di--Tevatron the
increased $L$ and/or $\sqrt s$
enhances $N_{\sigma}$ and so the  $H\to \tau\nu_{\tau}$ channel offers
the best prospects for Model I. In Model I$'$ this is the case even when $L=2$
fb$^{-1}$.

\begin{figure}[t]
\begin{center}
\mbox{\epsfig{file=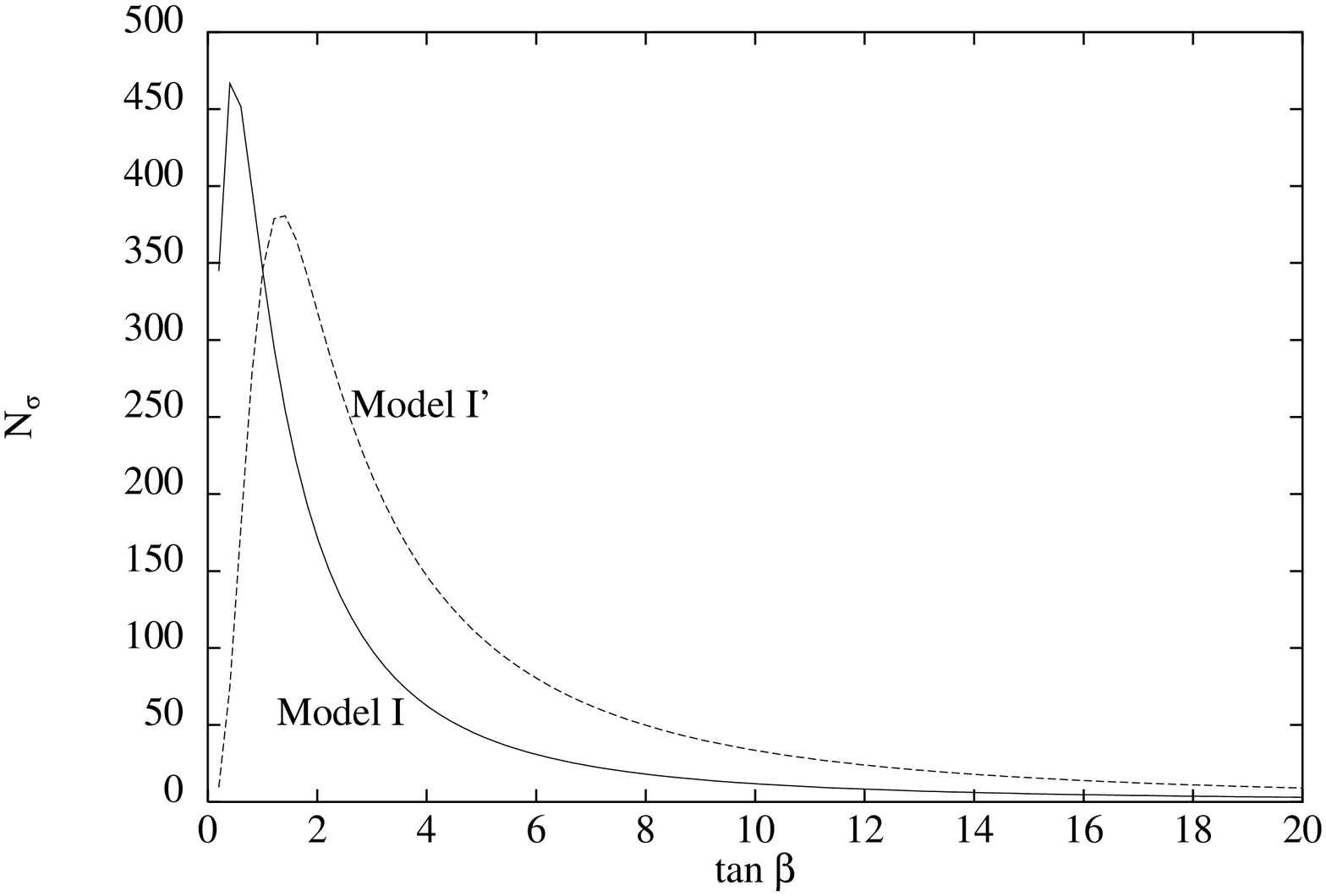,angle=-90,height=9cm}}
\end{center}
\vspace{-5mm}
\caption{$N_{\sigma}$ as a function of $\tan\beta$ for both Model I$'$
and Model I$'$ at the LHC. $M_H=130$ GeV is taken.}
\label{Fig:Top4}
\end{figure}

Finally we shall consider the LHC. Using the same values for the various
efficiencies as was used at the Tevatron, Figure \ref {Fig:Top4} shows
that for $M_H=130$ GeV in Model I$'$ (Model I) $N_{\sigma}\ge 5$ is
maintained until
$\tan\beta=25$ (15).
However, using similar efficiencies to those used in
Ref. \cite{Hubb} (which are for the SSC) we find that the above values of
$\tan\beta=25$ and 15 must be replaced by 8 and 5. A $b$--tag  is
included here and $e_l=0.46$ is taken, while at the Tevatron $e_l\approx 1.0$
and no $b$--tag is required. Also, Ref. \cite{Hubb} does not make use
of $\tau^-\to \rho^-\nu_{\tau}\to \pi^0\pi^-\nu_{\tau}$. Despite this
reduced coverage, it is
clear that the LHC will probe a significant region of $\tan\beta$ and $M_H$
parameter space, especially if the efficiencies used in Ref. \cite{Hubb}
are improved.

\section{Conclusions}
We have studied the prospects for detecting charged Higgs scalars $(H^{\pm})$
contributing to top quark decay ($t\to Hb$) in the context of the
non--minimal Standard Model. Considered were $H^{\pm}$s from the 2HDM
(Models I and I$'$), HTM and MHDM, all of which escape the mass bounds
from $b\to s\gamma$ and thus may be light enough to contribute to top
quark decay. Two detection methods were presented:
\begin{itemize}
\item[{(i)}] Searching for an excess of energetic pions ($\pi^{\pm}$) over
that predicted from lepton universality.
\item[{(ii)}] Reconstructing the quark decays of $H^{\pm}$.
\end{itemize}
Neither method is able to extract a statistically significant signature of
$H^{\pm}$ with
the current CDF data sample (67 pb$^{-1}$) at the Tevatron, and even the
expected 140 pb$^{-1}$ (available by the end of 1995) will prove
insufficient over most of the parameter ($M_H$, $\tan\beta$) space.
Thus the existence of a light $H^{\pm}$ (50 GeV $\le M_H \le m_t-m_b$) is
fully consistent with current data, and may be searched for at LEP2.
Prospects are much better at proposed higher energy/luminosity colliders.
At an
upgraded Tevatron of $L=2$ fb$^{-1}$ method
(i) will provide (in Model I$'$ for $M_H=80$ GeV) an energetic pion excess of
$5\sigma$  ($3\sigma$)
 if $\tan\beta\le 5.5$ (8). For $M_H=130$ GeV an excess of
$5\sigma$
($3\sigma$) is maintained until  $\tan\beta=3$ (5). For Model I the
situation is worse due to the inferior BR $(H\to \tau\nu_{\tau})$.
A Di--Tevatron with $\sqrt s=4$ TeV and $L=20$ fb$^{-1}$ would provide a
$5\sigma$ signal for $\tan\beta\le 11$ (8) in Model I$'$ (Model I) for
$M_H=130$ GeV. The LHC would improve the coverage to 25 and 15 respectively.

Method (ii) is necessary for the case of a leptophobic Higgs with BR ($H\to
{\rm jets})\approx 100\%$ which is possible in the MHDM. A noticeable
peak at $M_H$
can be obtained for $M_H=80$ GeV (130 GeV) if $\tan\beta\le 3.5$ (2).
For larger $\tan\beta$ the detection of a leptophobic $H^{\pm}$ appears
unlikely in top quark decay and
other production methods must be considered. For Model I method (ii) is
competitive
with method (i) at the Tevatron ($L=2$ fb$^{-1}$), but in all other
cases method (i) offers the best detection prospects.

\section*{Acknowledgements}
I wish to thank W.J. Stirling and D.P. Roy for useful comments. This
work has been supported by the UK EPSRC research council.

\end{document}